\documentclass[prb,aps,twocolumn,superscriptaddress]{revtex4-2}
\usepackage{amsmath}
\usepackage{color}
\usepackage{bbm}
\usepackage{amssymb}
\usepackage{epsfig}
\usepackage{multirow}
\usepackage{amsbsy}
\usepackage{array}
\usepackage{diagbox}
\usepackage{bm}
\usepackage{extarrows}
\usepackage{graphicx}
\usepackage{subfigure}
\usepackage{appendix}
\usepackage{txfonts}
\usepackage{lipsum}
\usepackage{bbding}
\usepackage{pifont}
\usepackage{makecell}
\usepackage[version=4]{mhchem}
\allowdisplaybreaks[4]
\usepackage[colorlinks=true,linkcolor=blue,citecolor=blue,urlcolor=blue,bookmarks=false]{hyperref}

\begin{document}

\title{Interband-Pairing–Boosted Supercurrent Diode Effect in Multiband Superconductors}

\author{Jiong Mei}
\affiliation{Beijing National Laboratory for Condensed Matter Physics and Institute of Physics, Chinese Academy of Sciences, Beijing 100190, China}
\affiliation{School of Physical Sciences, University of Chinese Academy of Sciences, Beijing 100190, China}

% \author{Kun Jiang}
% \affiliation{Beijing National Laboratory for Condensed Matter Physics and Institute of Physics, Chinese Academy of Sciences, Beijing 100190, China}
% \affiliation{School of Physical Sciences, University of Chinese Academy of Sciences, Beijing 100190, China}

\author{Shengshan Qin}\email{qinshengshan@bit.edu.cn}
\affiliation{School of Physics, Beijing Institute of Technology, Beijing 100081, China}

\author{Jiangping Hu}
\email{jphu@iphy.ac.cn}
\affiliation{Beijing National Laboratory for Condensed Matter Physics and Institute of Physics, Chinese Academy of Sciences, Beijing 100190, China}
\affiliation{Kavli Institute for Theoretical Sciences and CAS Center for Excellence in Topological Quantum Computation, University of Chinese Academy of Sciences, Beijing 100190, China}
\affiliation{New Cornerstone Science Laboratory, Beijing, 100190, China}

\begin{abstract}
% We reveal a new mechanism realizing strong supercurrent diode effect in Josephson junctions constructed by multiband superconductors. In a multiband superconductor whose energy bands are nearly degenerate, Cooper pair can form between two electrons in the same band or in different bands. We find that the interband pairing can greatly enhance the supercurrent diode effect. Impressively, considerable diode efficiency can be achieved even in the weak spin-orbit coupling condition, where it is vanishing in the intraband pairing condition. We demonstrate the mechanism analytically and numerically. We apply the theory to the monolayer FeSe/STO, where evidences for interband pairing have been observed in recent experiments and the interband pairing may be attributed to the nodeless $d$-wave pairing or the $\eta$ pairing states. With parameters fit from experiments, we predict the monolayer FeSe/STO a potential high-temperature platform supporting large supercurrent diode effect with the diode efficiency up to $30\%$ ($12\%$) in the $d$-wave ($\eta$) pairing state. Our study also suggests the measurements of the supercurrent diode effect can provide essential information for the pairing state in the monolayer FeSe/STO.
We unveil a mechanism that enables a robust supercurrent diode effect in Josephson junctions based on multiband superconductors. We predict that interband pairing can significantly amplifies this effect, even under weak spin-orbit coupling while intraband pairing alone would render it negligible. To illustrate this, we examine monolayer FeSe/STO, a system where recent experiments suggest interband pairing in either a nodeless $d$-wave or $\eta$ pairing state. Using experimentally derived parameters, we predict that FeSe/STO can serve as a high-temperature platform for realizing a substantial supercurrent diode effect, with efficiencies reaching up to $30\%$ for $d$-wave pairing and $12\%$ for $\eta$ pairing. These results demonstrate that measuring the supercurrent diode effect can provides a powerful probe of the pairing symmetry in monolayer FeSe/STO, offering critical insights into its superconducting state.
\end{abstract}

\maketitle

\textit{Introduction.}
The Josephson diode effect (JDE) is a nonreciprocal charge transport phenomenon in Josephson junctions\cite{JOSEPHSON1962251,PhysRevLett.10.230,PhysRevLett.99.067004}. It is characterized by different critical supercurrents in the forward and reversal directions in the junction, i.e. $|I_{c+}|\neq|I_{c-}|$, and its realization is always accompanied by the breaking of the time-reversal symmetry and the inversion symmetry. The JDE can lead to a unidirectional nondissipative charge transport can be realized in Josephson junctions, which have promising applications in superconductor electronics. In recent years, research on the Josephson diode effect has advanced rapidly, with numerous theoretical proposals emerging\cite{PhysRevB.98.075430,PhysRevB.103.245302,10.1126/sciadv.abo0309,PhysRevX.12.041013,PhysRevLett.129.267702,PhysRevB.106.134514,PhysRevLett.130.266003,PhysRevLett.131.096001,PhysRevLett.130.177002,PhysRevB.107.245415,PhysRevB.108.214519,PhysRevB.108.054522,PhysRevResearch.5.033199,Karabassov2024,PhysRevB.109.L081405,PhysRevB.109.174511,PhysRevB.110.155405,PhysRevB.110.014518,guo2024phi0junction,Debnath_2025,patil2024andreev,PhysRevB.111.L140506,PhysRevB.111.174515,10.21468/SciPostPhys.17.2.037}. According to the present studies, to obtain a large JDE efficiency which is vital for applications, usually a large spin-orbit coupling is needed. Significant efforts have been made experimentally, resulting in considerable progress\cite{Ando2020,Pal2022,Wu2022,Baumgartner_2022,Turini2022,Trahms2023,DM2023,PhysRevResearch.5.033131,Ghosh2024,Reinhardt2024,Li2024NC,PhysRevB.110.104510, qi2025high, zeng2025large}. The JDE has been successfully realized in various systems, including Josephson junctions (JJs) formed by Dirac semimetal $\mathrm{NiTe_2}$ coupled with superconducting Nb electrodes\cite{Pal2022}, van der Waals  $\mathrm{NbSe_2/Nb_3Br_8/NbSe_2}$ heterostructures\cite{Wu2022}, arrays of JJs fabricated on epitaxial Al/InAs heterostructures\cite{Baumgartner_2022}, the twisted $\mathrm{Bi_2Sr_2CaCu_2O_{8+\delta}}$\cite{Ghosh2024} and the twisted $\mathrm{FeTe_{0.55}Se_{0.45}/NbSe_2}$\cite{zeng2025large}.

On the other hand, among the discovered superconductors a large number are multiband superconductors. The multiband effects can lead to exotic phenomena in superconductors, such as the topological superconductivity\cite{RevModPhys.83.1057,RevModPhys.88.035005}. However, in the current studies the multiband effect on the JDE has been rarely involved. In this work, we study the JDE in Josephson junctions constructed by multiband superconductors. In multiband superconductors, Cooper pairs can form between electrons in the same band or in different bands, when the normal-state energy bands are nearly degenerate. We find that the interband pairing can substantially enhance the diode efficiency and strong JDE can be realized even in the weak spin-orbit coupling limit, compared to the intraband pairing condition. We reveal that such enhancement arises from the profound effects of the interband pairing, the spin-orbit coupling and the Zeeman field on the Andreev bound states in the junction. We apply the analysis to the monolayer FeSe/STO, which is a typical multiband supercondtor with signatures for interband pairing observed in recent experiments\cite{ding2024sublatticedichotomymonolayerfese}. The interband pairing component in the monolayer FeSe/STO may be attributed to the nodeless $d$-wave pairing\cite{Hirschfeld_2011,RevModPhys.85.849,annurev:/content/journals/10.1146/annurev-conmatphys-031016-025242} or the $\eta$ pairing\cite{PhysRevX.3.031004}, candidate pairings proposed in previous theoretical studies. Based on a thorough analysis, we predict large diode efficiency in both cases and suggest the monolayer FeSe/STO a potential high-temperature platform supporting strong JDE.

\begin{figure}
    \centering
    \epsfig{figure=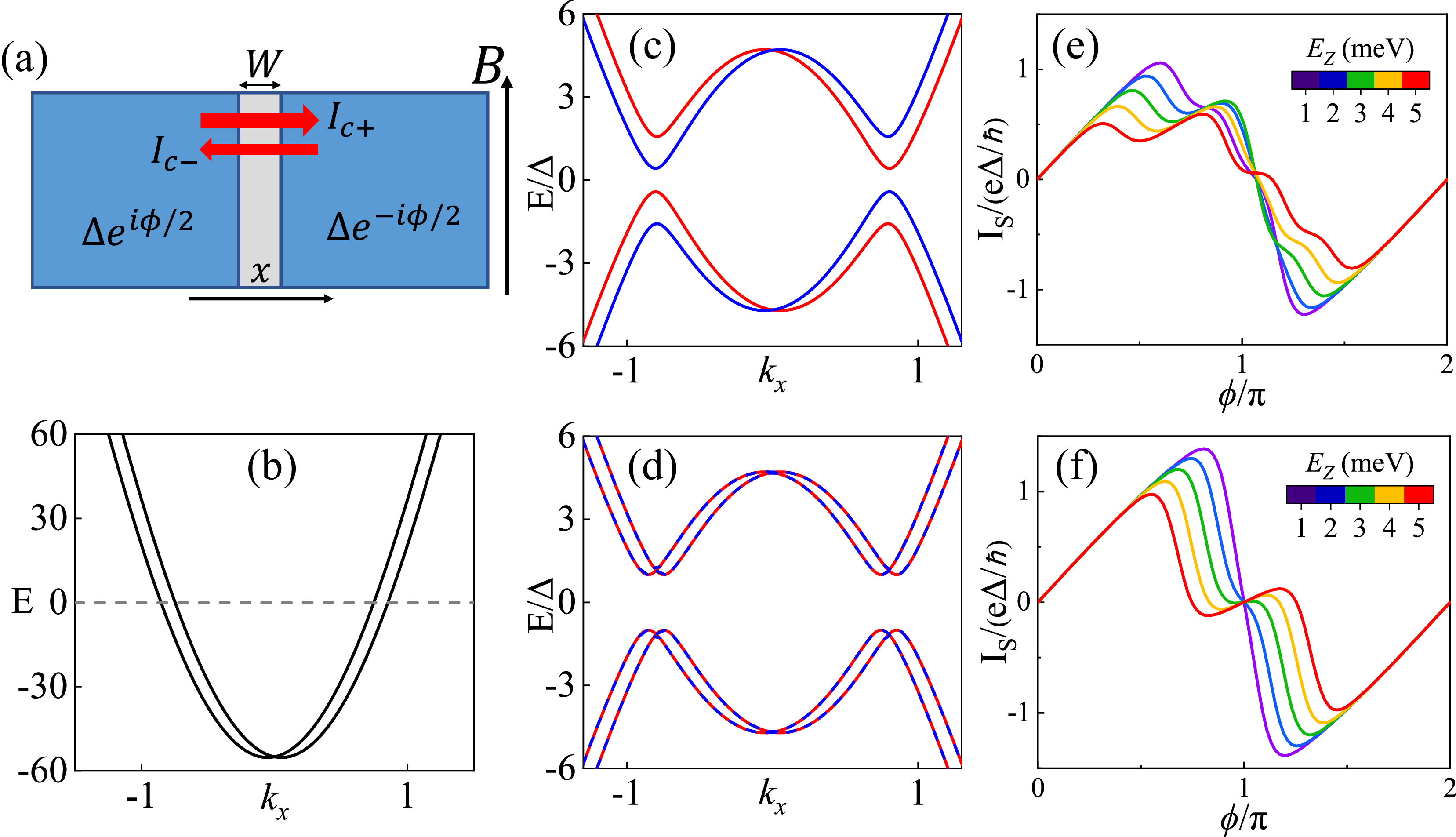, width=0.49\textwidth}
    \caption{(a) sketches the 1D JJ. A magnetic field $\bm{B}$ is applied perpendicular to the system. The gray region is metallic with a width $W$. (b) shows the normal-state band structure of $h({\bm k})$ in Eq.\eqref{eq_1D} with all bands being doubly degenerate. (c) and (d) present the energy spectra for the superconductor leads in $h({\bm k})$ in the pure interband pairing ($\Delta_{intra}=0$, $\Delta_{inter}=12$) and pure intraband pairing ($\Delta_{intra}=12$, $\Delta_{inter}=0$) conditions respectively. The red and blue lines represent the energy spectra from the two decoupled subsystems as illustrated in Eq.\eqref{eq_1Dwhole_band}. (e) and (f) show the CPR for the Josephson current corresponding to the pure interband pairing and pure intraband pairing cases in (c) and (d) respectively. The CPR are calculated for different magnetic fields $\bm{B}$, with a fixed inversion-symmetry breaking perturbation $\delta\mu=-1$ in Eq.\eqref{eq_1Dbreak}. In simulating the figures, the other parameters are taken as $\{\lambda_0,\lambda_1,\mu\}=\{90.8, 3.86, 55\}$;  In (b)$\sim$(d), the SOC is enlarged by 2.5 times to show the spectra clearer.}
    \label{fig1}
\end{figure}

\textit{1D toy model.}
We first use a 1D JJ sketched in Fig.\ref{fig1}(a) to demonstrate the mechanism. The junction is comprised of two superconductor leads with a phase difference $\phi$, and the normal-state part of the superconductor leads is a two-orbital electron gas. To describe the junction, we introduce the following Hamiltonian
\begin{align}\label{eq_1D}
h({\bm k}) &= \lambda_0k_{x}^{2}\tau_3 + \lambda_1 k_{x}\sigma_{3}s_{2}\tau_3 - \mu \tau_3  \nonumber\\
&+ f(\bm{r})\tau_{+}\Delta(\bm{k})+f^{*}(\bm{r})\tau_{-}\Delta^{\dagger}(\bm{k}),
\end{align}
where $\sigma$, $s$ and $\tau$ are the Pauli matrices representing the orbital, spin and Nambu degrees of freedom respectively, and $\tau_{\pm}=(\tau_1\pm i\tau_2)/2$. The Hamiltonian in Eq.\eqref{eq_1D} is written in the basis $(c_{\bm{k}},\,c_{-\bm{k}}^{\dagger})^{T}$, where the indices for the spin and orbital degrees are omitted for simplicity. In $h({\bm k})$, the first term is the kinetic energy, the second term is the Rashba spin-orbit coupling, the third term describes the chemical potential, and the last two terms correspond to the superconducting order. In the junction, the function $f(\bm{r})$ in the superconducting part takes the form $f(\bm{r})=e^{-i \mathrm{sgn}(x)\phi/2}\Theta(|x-W/2|-W/2)$, with $\mathrm{sgn}()$ being the sign function, $\Theta()$ the step function, and $W$ the width of the junction. We note that in the system in Eq.\eqref{eq_1D}, the normal-state Hamiltonian in $h({\bm k})$ respects both the inversion symmetry $\mathcal{I}$ and the time-reversal symmetry $\mathcal{T}$, and the two operators take the matrix form $\mathcal{I}=\sigma_1$ and $\mathcal{T}=is_2K$ with $K$ the complex conjugation operation.

For the superconductivity in Eq.\eqref{eq_1D}, we focus on the intraorbital spin-singlet pairing and the interorbital spin-singlet pairing, namely
\begin{align}\label{eq_1Dpair}
\Delta(\bm{k}) = \Delta_{intra}is_2 + \Delta_{inter}\sigma_1is_2.
\end{align}
By applying a rotation $\mathcal{R} = e^{is_1\pi/4}$ to the normal-state Hamiltonian in Eq.\eqref{eq_1D}, we can project the superconducting orders in Eq.\eqref{eq_1Dpair} onto the band basis. In the band basis, one can find that the intraorbital pairing $\Delta_{intra}$ is purely intraband, while the interorbital pairing $\Delta_{inter}$ is a purely interband. Here, it is worth pointing out that usually the interband pairing order is not favored; However, when the Fermi surfaces from different bands are nearly degenerate as indicated in Fig.\ref{fig1}(b), the interband pairing or at least a interband component becomes possible. The superconducting energy spectra in the interorbital (interband) and intraorbital (intraband) cases are illustrated in Figs.\ref{fig1}(c)(d).

To induce the JDE, the inversion and the time-reversal symmetries in the junction need to be broken simultaneously. We introduce the following terms into the junction in Eq.\eqref{eq_1D}
\begin{equation}\label{eq_1Dbreak}
h_{br}({\bm k}) = \delta\mu\sigma_{3}\tau_3 + E_{Z}s_{2},
\end{equation}
where $\delta\mu$ is a perturbation breaking the inversion symmetry, and $E_Z$ is the Zeeman coupling introduced by an external magnetic field applied perpendicular to the system as sketched in Fig.\ref{fig1}(a).

\textit{JDE in 1D.}
We can solve the JJ depicted by $h({\bm k}) + h_{br}({\bm k})$ analytically in the short junction condition. Here, we focus on the pure interband pairing condition, i.e. $\Delta_{intra} = 0$. In the band basis, the system can be decoupled as
\begin{align}\label{eq_1Dwhole_band}
H_{BdG}^{inter}({\bm k}) & =\oplus_{i}H_{BdG}^{inter,i}({\bm k}), \quad i=1,2
\end{align}
with
\begin{align}
H_{BdG}^{inter,i}({\bm k}) & =\begin{pmatrix}H_{inter,i}({\bm k})-\mu & \Delta is_{2}\\
-\Delta^\ast is_{2} & \mu-H_{inter,i}^{*}(-{\bm k})
\end{pmatrix}.
\end{align}
In the above equation, $H_{inter,1}({\bm k}) =(\lambda_{0}k_{x}^{2}-\lambda_{1}k_{x})s_{0}+(E_{Z}-\delta\mu)s_{3}$ and $H_{inter,2}({\bm k}) =(\lambda_{0}k_{x}^{2}+\lambda_{1}k_{x})s_{0}+(E_{Z}+\delta\mu)s_{3}$.
Based on the scattering matrix formalism\cite{PhysRevLett.67.3836,10.1126/sciadv.abo0309,10.1007/978-3-642-84818-6_22}, we calculate the Andreev bound states contributed by $H_{BdG}^{inter,1}(k)$. These Andreev bound states take eigenvalues $\pm E_{1,1}$ and $\pm E_{1,2}$ with
\begin{align}\label{eq_ABSH1}
E_{1,j} = \Delta\cos\left(\frac{\tilde{\phi}_{1,j}}{2}\right)- \tilde{\Delta}_{1,j}, \quad j=1,2
\end{align}
where
\begin{align}\label{eq_ABSH12}
\tilde{\phi}_{1,j}=\phi+\frac{\pi\tilde{\Delta}_{1,j}}{E_{T}(1-\lambda_{1}^{2}/v_{F}^{2})}, \  \tilde{\Delta}_{1,j}=\lambda_{1}k_{F}+(-1)^{j}(\delta\mu-E_{Z}).
\end{align}
The spectra of the Andreev bound states attributed to $H_{BdG}^{inter,2}(k)$ are $\pm E_{2,1}$ and $\pm E_{2,2}$ with
\begin{align}\label{eq_ABSH2}
E_{2,j} & =\Delta\cos\left(\frac{\tilde{\phi}_{2,j}}{2}\right)-\tilde{\Delta}_{2,j}, \quad j=1,2
\end{align}
where
\begin{align}\label{eq_ABSH22}
\tilde{\phi}_{2,j}=\phi+\frac{\pi\tilde{\Delta}_{2,j}}{E_{T}(1-\lambda_{1}^{2}/v_{F}^{2})}, \  \tilde{\Delta}_{2,j}=-\lambda_{1}k_{F}+(-1)^{j}(\delta\mu+E_{Z}).
\end{align}
Here, $k_F$ is the Fermi momentum satisfying $\lambda_0 k_F^2 = \mu$, $v_F=2\lambda_0k_F$ is the Fermi velocity, and $E_T=(\pi/2)v_F/W$ is the corresponding Thouless energy.

% \begin{figure}
%     \centering
%     \epsfig{figure=ABS, width=0.36\textwidth}
%     \caption{The ABS spectrum for Hamiltonians $H_{BdG}^{inter,1}$ (red lines) and $H_{BdG}^{inter,2}$ (blue lines). The peak of positive (negative) CPR is indicated by the cyan (green) vertical lines and the gray dashed line indicates the position of $\phi=\pi$. The parameters are chosen as $\{\lambda_0/a^2,\lambda_1/a,\delta\mu,E_Z,\Delta,\mu\}=\{90.8,3.86,-1,4,12,55\}$.}
%     \label{fig2}
% \end{figure}

\begin{figure}
    \centering
    \epsfig{figure=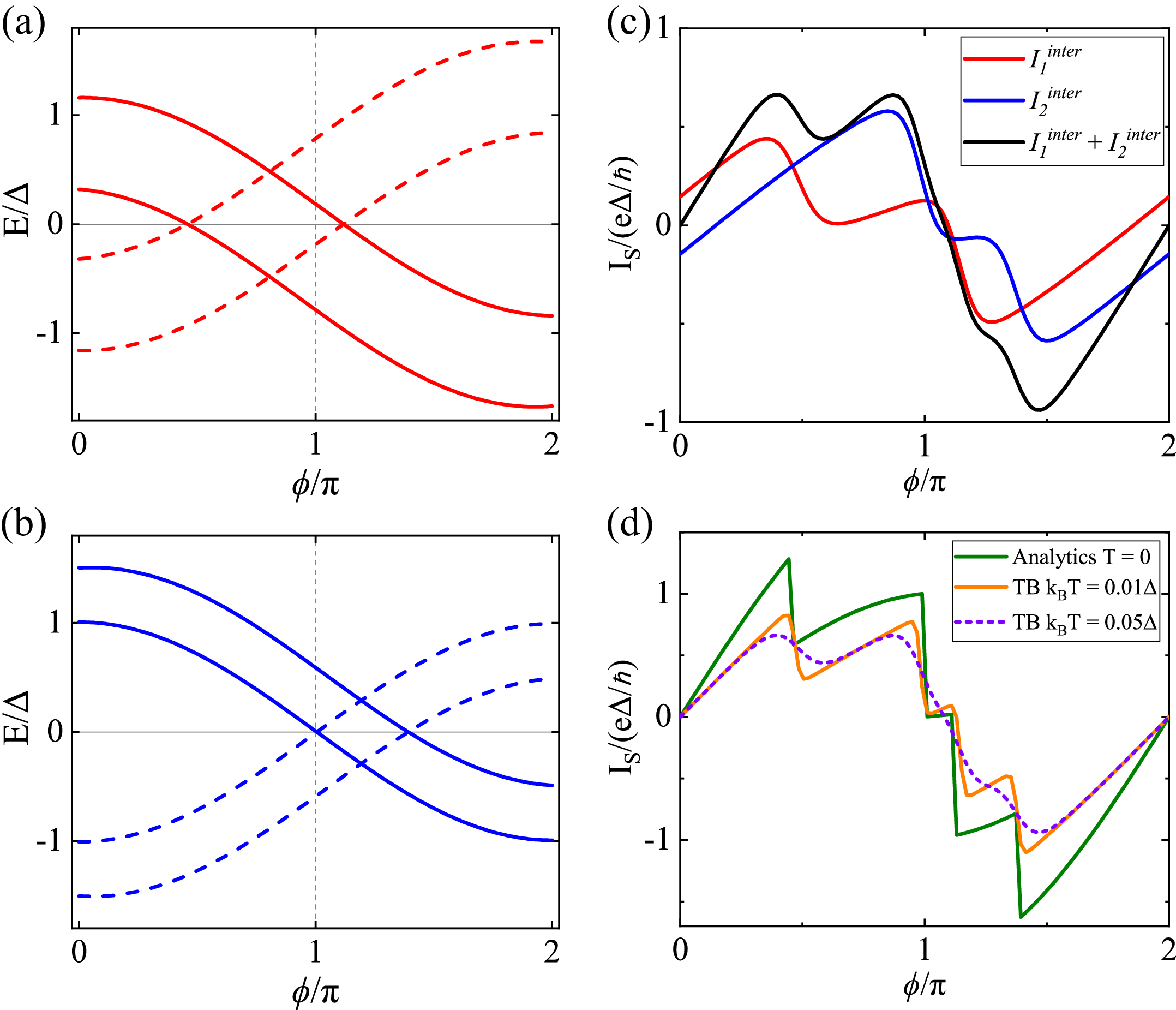, width=0.45\textwidth}
    \caption{(a) and (b) show the spectrum for the Andreev bound states in the 1D junction in Fig.\ref{fig1}, with the contributions from $H_{BdG}^{inter,1}$ and $H_{BdG}^{inter,2}$ in Eq.\eqref{eq_1Dwhole_band} presented in (a) and (b) respectively. The right (left)-moving modes are marked with the dashed (solid) lines. (c) shows the total Josephson current (black) in the junction, with the contributions from $H_{inter,1}$ and $H_{inter,2}$ marked in red and blue respectively. The calculations are done at $k_BT=0.05\Delta$. (d) presents the total Josephson current in the junction calculated from the analytical analysis (green) and tight-binding simulations (orange and violet). In the calculations, we set $E_Z = 4$ and $\Delta_{inter} = 12$ and the other parameters are the same with those in Fig.\ref{fig1}.}
    \label{fig2}
\end{figure}

The Josephson current in the junction can be calculated as $I(\phi)=\frac{2e}{\hbar}\frac{\mathrm{d}F}{\mathrm{d}\phi}$. Here, $F$ is the free energy of the whole system, including the contributions from the Andreev bound states and the continuum states in the bulk. Generally, both the Andreev bound states and the continuum states contribute to the supercurrent in the junction. However, in the interband pairing condition we find that the contributions of the continuum states from $H_{BdG}^{inter,1}({\bm k})$ and $H_{BdG}^{inter,2}({\bm k})$ compensate with each other, and only the contributions of the Andreev bound states survive. At zero temperature, the supercurrent in the junction can be expressed as $I_S = I_1^{inter} + I_2^{inter}$, where
\begin{align}\label{eqn10}
I_{1}^{inter}(\phi) & =\frac{e\Delta}{2\hbar}\sum_{j=1,2}\sin\frac{\tilde{\phi}_{1,j}}{2}\mathrm{sgn}\left(\cos\frac{\tilde{\phi}_{1,j}}{2}-\frac{\tilde{\Delta}_{1,j}}{\Delta}\right)+\frac{2e\lambda_{1}k_{F}}{\pi\hbar}\nonumber \\
I_{2}^{inter}(\phi) & =\frac{e\Delta}{2\hbar}\sum_{j=1,2}\sin\frac{\tilde{\phi}_{2,j}}{2}\mathrm{sgn}\left(\cos\frac{\tilde{\phi}_{2,j}}{2}-\frac{\tilde{\Delta}_{2,j}}{\Delta}\right)-\frac{2e\lambda_{1}k_{F}}{\pi\hbar}.
\end{align}
% and
% \begin{align}\label{eqn11}
% \tilde{\phi}_{1,\pm} & = \phi+\frac{\pi\tilde{\Delta}_{1,\pm}}{E_{T}(1-\lambda_{1}^{2}/v_{F}^{2})}, \ 
% \tilde{\Delta}_{1,\pm} = \lambda_{1}k_{F} \pm (\delta\mu-E_{Z}), \nonumber\\
% \tilde{\phi}_{2,\pm} & = \phi+\frac{\pi\tilde{\Delta}_{2,\pm}}{E_{T}(1-\lambda_{1}^{2}/v_{F}^{2})}, \ 
% \tilde{\Delta}_{2,\pm} = -\lambda_{1}k_{F} \pm (\delta\mu+E_{Z}),
% \end{align}
% with $E_T=(\pi/2)v_F/W$ being the Thouless energy.
In Eq.\eqref{eqn10}, $I_{1}^{inter}$ and $I_{2}^{inter}$ are the Josephson currents contributed by $H_{BdG}^{inter,1}({\bm k})$ and $H_{BdG}^{inter,2}({\bm k})$ respectively; And in both $I_{1}^{inter}$ and $I_{2}^{inter}$, the term proportional to $\Delta$ is attributed to the Andreev bound states, and the remaining constant term arises from the continuum states. We present more details on the calculations of the Andreev bound states and the Josephson current in the SM\cite{SuppMat}.

Based on the above analytical results, we plot the spectra of the Andreev bound states in Figs.\ref{fig2}(a)(b), and the Josephson currents are shown in Fig.\ref{fig2}(d). The current-phase relation (CPR) in Fig.\ref{fig2}(d) reveals a pronounced JDE in the junction. We also simulate the CPR based on the lattice model transformed from the continuum model $h({\bm k}) + h_{br}({\bm k})$. The results are presented in Fig.\ref{fig2}(c), which agree well with the analytical results. The results in Figs.\ref{fig2}(c)(d) indicates that in the interband pairing case, a weak spin-orbit coupling can lead to a considerable supercurrent diode efficiency $\gamma=\frac{I_{c,+}-|I_{c,-}|}{I_{c,+}+|I_{c,-}|}$; And it reaches up to $\gamma = 17\%$ in the tight-binding simulation shown in Fig.\ref{fig2}(c). To see this clearer, we compare the JDE in the junction in the pure intraband pairing condition and in the pure interband pairing condition. The results are presented in Figs.\ref{fig1}(e)(f), which show the vanishing diode efficiency in the intraband pairing case while remarkable efficiency in the interband pairing case.

To understand the boosting effect of the interband pairing on the JDE, we go back to the analytical formula for the Andreev bound states in Eq.\eqref{eq_ABSH1}\eqref{eq_ABSH2} and Josephson current in Eq.\eqref{eqn10}. As mentioned, in the interband pairing case the supercurrent is thoroughly contributed by the Andreev bound states, which can be straightforwardly conjectured from the slope of the positive-energy Andreev bound states. In the short junction limit, we have $\tilde{\Delta}_{1(2),j}\ll E_T$, namely $\tilde{\phi}_{1(2),j}\approx\phi$. Therefore, the critical supercurrent is mainly determined by the factor $\mathrm{sgn}\left(\cos(\phi/2)-\tilde{\Delta}_{1(2),j}/\Delta\right)$ in Eq.\eqref{eqn10}. Based on Eqs.\eqref{eq_ABSH1}\eqref{eq_ABSH2}\eqref{eqn10}, it can be inferred once the energy of the Andreev bound states changes sign, the supercurrent reaches a local maximum as indicated in Fig.\ref{fig2}; And the critical supercurrent in each direction corresponds to the largest maximum in the direction. According to the above analysis, we conclude two major reasons for the boosting effect. Firstly, in the interband pairing case the energy of the Andreev bound states shifts linearly as the spin-orbit coupling $\lambda_1$ varies as shown in Eqs.\eqref{eq_ABSH1}\eqref{eq_ABSH2}, while in the intraband pairing case the spin-orbit coupling affects the Andreev bound states in a more minor way (details in SM\cite{SuppMat}). The $\delta\mu$ and $E_Z$ terms play similar roles with $\lambda_1$ pairing condition as shown in Eqs.\eqref{eq_ABSH1}\eqref{eq_ABSH2}. This makes the JDE more sensitive to the perturbations in the interband pairing condition. Secondly, the multiband effect leads to multiple branches of the Andreev bound states as shown in Figs.\ref{fig2}(a)(b) and multiple supercurrent peaks in $I^{inter}_1$ and $I^{inter}_2$ in the CPR shown in Fig.\ref{fig2}(c). We consider the diode effect attributed to $I^{inter}_1$ and $I^{inter}_2$ separately, defining $\gamma_i = \frac{ I^{c,+}_i - |I^{c,-}_i|}{ I^{c,+}_i + |I^{c,-}_i|}$ with $I^{c,+}_i$ and $I^{c,-}_i$ the largest supercurrent in opposite directions in $I^{inter}_i$. As shown in Fig.\ref{fig2}(c), $I^{c,-}_1$ and $I^{c,-}_2$ in the region $\pi\leq\phi\leq2\pi$ locate closer than $I^{c,+}_1$ and $I^{c,+}_2$ in the region $0\leq\phi\leq\pi$. Such misalignment of the supercurrent peaks in $I^{inter}_1$ and $I^{inter}_2$ leads to $\gamma > \gamma_i$, further enhancing the JDE in the junction.

In the above, we show strong JDE can be realized in the weak spin-orbit coupling condition in the interband pairing case. However, a zero spin-orbit coupling will lead to the vanishing diode efficiency in the junction. According to Eqs.\eqref{eq_ABSH12}\eqref{eq_ABSH22}, When $\lambda_1 = 0$ we have $\tilde{\Delta}_{1(2),j=1}=-\tilde{\Delta}_{1(2),j=2}$. This makes the critical current in each $I^{inter}_i$ in Eq.\eqref{eqn10} symmetric around $\phi=\pi$, resulting in the vanishing $\gamma_i$ and $\gamma$. Another interesting point in the model is that in the condition $E_z = \delta\mu = 0$, $\gamma_1$ and $\gamma_2$ can be nonzero because in each of the subsystems neither the time reversal symmetry nor the inversion symmetry exits. However, $\gamma_1$ and $\gamma_2$ compensate with each other, i.e. $\gamma_1 = -\gamma_2$, due to the fact that the time-reversal symmetry and the inversion symmetry map the two subsystems to each other. Similar phenomenon occurs in the condition $E_z = 0$ or $\delta\mu=0$.

% \begin{figure}
%     \centering
%     \subfigure[]{
%     \epsfig{figure=1d_TB_inter, width=0.23\textwidth}
%     }
%     \subfigure[]{
%     \epsfig{figure=1d_TB_intra, width=0.23\textwidth}
%     }

%     \subfigure[]{
%     \epsfig{figure=case2_H1H2, width=0.24\textwidth}
%     }
%     \subfigure[]{
%     \epsfig{figure=analytics_TB, width=0.22\textwidth}
%     }
%     \caption{[(a) and (b)] The Josephson CPR $I_S$ for in-plane Zeeman field $E_Z$ going from $1\,meV$ to $5\,meV$ for interband pairing and intraband pairing, respectively. The Josephson current is calculated from the tight binding model of Hamiltonian~\eqref{eqn1} by setting the lattice constant $a=1$, where the other parameters are chosen as $\{\lambda_0,\lambda_1,\delta\mu,\Delta,\mu\}=\{90.8,3.86,-1.0,12,55\}$ in units of $meV$. (c) The contributions of Josephson current from two sets of Hamiltonian $H_{inter,1}$ and $H_{inter,2}$ for $E_Z=4\,meV$. (d) The analytical (green solid) and tight binding (orange solid and violet dashed) calculations of $I_S$ for $E_Z=4\,meV$. The temperature is set to be $k_BT=0.05\Delta$ in [(a), (b) and (c)]. }
%     \label{fig3}
% \end{figure}

\textit{JDE in monolayer FeSe/STO.}
The iron-based superconductors are typical multiband superconductors, among which the monolayer FeSe/STO has the highest superconducting transition temperature above $56 K$. However, consensus on its pairing symmetry has not been reached\cite{Hirschfeld_2011,RevModPhys.85.849,annurev:/content/journals/10.1146/annurev-conmatphys-031016-025242}. Recently, scanning tunneling spectroscopy measurements reveal a sublattice dichotomy in the superconducting coherence peak in the monolayer FeSe/STO\cite{ding2024sublatticedichotomymonolayerfese}. Such a phenomenon has two important implications. Firstly, the inequivalence of the two Fe sublattices signals unambiguous inversion-symmetry breaking. It is worth mentioning that two other recent experimental works\cite{Wei_2025,zhang2024visualizinguniformlatticescalepair} also observe evidences for the inversion-symmetry breaking in the FeSe/STO. Secondly, the sublattice dichotomy indicates strong interband pairing in the monolayer FeSe\cite{ding2024sublatticedichotomymonolayerfese, PhysRevB.110.094517, to.appear}. Based on a thorough symmetry analysis\cite{to.appear}, the interband pairing component is most likely attributed to the nodeless $d$-wave pairing\cite{Hirschfeld_2011,RevModPhys.85.849,annurev:/content/journals/10.1146/annurev-conmatphys-031016-025242} or the $\eta$ pairing\cite{PhysRevX.3.031004}. We show the interband pairing, no matter it arising from the $d$-wave pairing or the $\eta$ pairing, can lead to significant JDE in the monolayer FeSe/STO. In the following in the main text, we focus on the $d$-wave condition and present the $\eta$ pairing condition in the SM\cite{SuppMat}.

%In recent studies, the authors conducted a systematic investigation of atomically flat monolayer FeSe using high-resolution STM/STS. Both studies find the tunneling spectra of $A$-Fe and $B$-Fe (shown in Fig.~\ref{fig3}(a)) are significantly different, resulting in the sublattice dichotomy. Since the two sublattices are related by inversion symmetry, these experimental results highlight the inversion-breaking nature of the superconducting state in monolayer FeSe, which satisfies a key requirement for the diode effect. Motivated by these experiments, we perform a detailed study of the diode effect in JJ formed by monolayer FeSe as schematically depicted in Fig.~\ref{fig1}.

\begin{figure}
    \centering
    \epsfig{figure=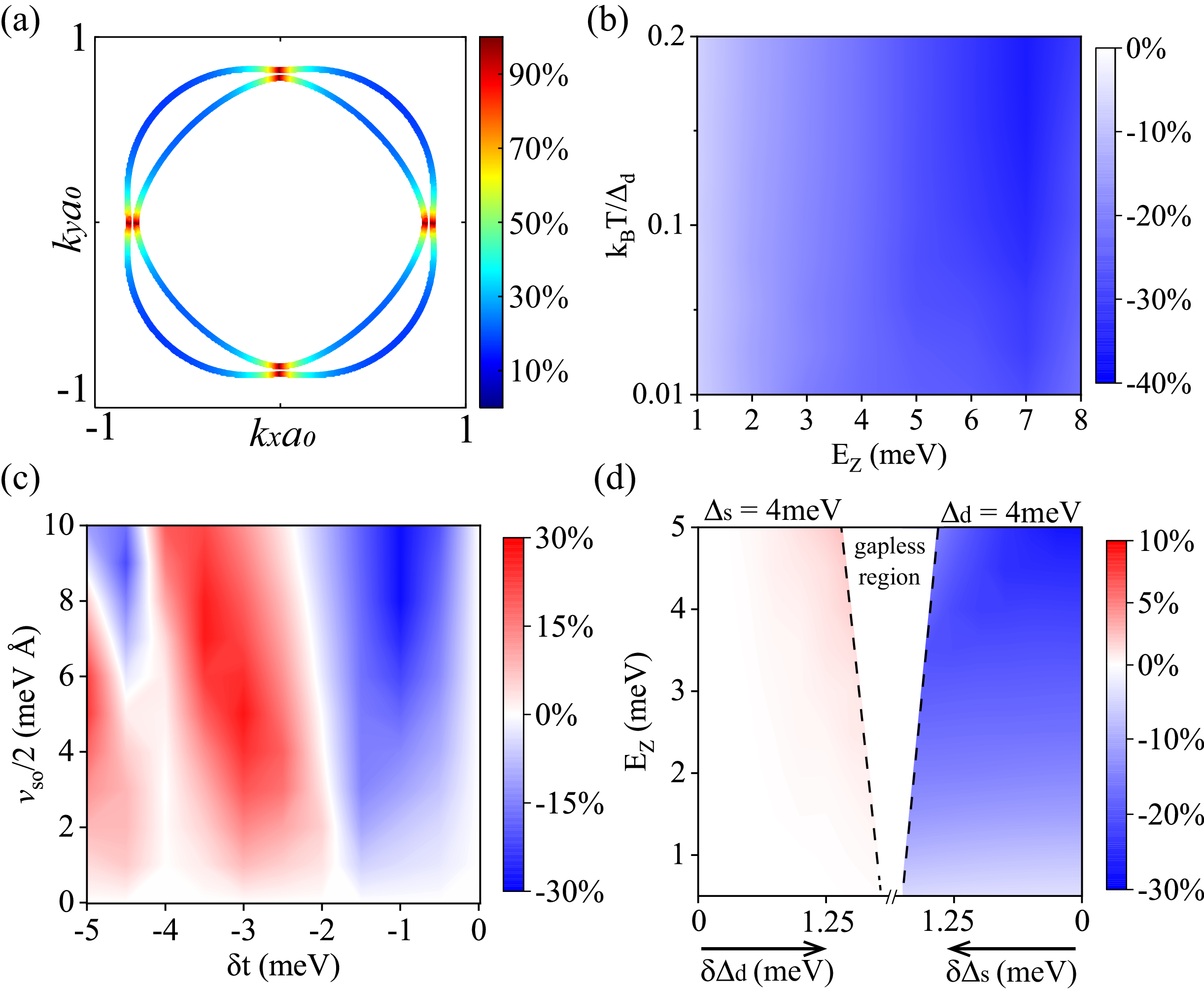, width=0.45\textwidth}
    \caption{(a) presents the ratio of the interband pairing component on the Fermi surfaces in the monolayer FeSe in the nodeless $d$-wave pairing case, where the colorbar represents $|\Delta_{inter}|/(|\Delta_{intra}|+|\Delta_{inter}|)$. (b)$\sim$(d) show the JDE efficiencies in the JJ constructed by the monolayer FeSe in different conditions. In (b) and (c), pure nodeless $d$-wave pairing is assumed with $\Delta_d=4\,\mathrm{meV}$; And (b) is calculated under a inversion-symmetry breaking perturbation $\delta t=-1\,\mathrm{meV}$, and (c) is under $E_Z = 5\,\mathrm{meV}$ and $k_BT = 0.1\Delta_d$. In (d), we assume the coexistence of the $d$-wave pairing and the extended $s$-wave pairing (the next-nearest-neighbour pairing) in the monolayer FeSe. In the left (right) side of the heatmap, the $s$ ($d$)-wave order dominates with fixed $\Delta_s=4\,\mathrm{meV}$ ($\Delta_d=4\,\mathrm{meV}$) and the mixed component $\delta\Delta_d$ ($\delta\Delta_s$) increases along the arrow directions; And we ignore the middle region where the superconducting state becomes nodal.}
    \label{fig3}
\end{figure}

To describe the normal state of the monolayer FeSe, we adopt the following effective model\cite{PhysRevLett.119.267001,qin2022spintriplet}
\begin{align}\label{eqn13}
h_{eff}({\bm k}) & = \frac{k^{2}}{2m} - \mu + v_{so}k_{x}\sigma_{3}s_{2} + v_{so}k_{y}\sigma_{3}s_{1} + \alpha k_{x}k_{y}\sigma_{1},
\end{align}
where $k^2 = k_x^2 + k_y^2$ is measured from $(\pi, \pi)$, and $\sigma$ and $s$ are Pauli matrices representing the orbital and spin degrees of freedom respectively. The model Hamiltonian $h_{eff}({\bm k})$ respects the time reversal symmetry $\mathcal{T} = is_2K$ and the $P4/nmm$ space group with the generators being the inversion symmetry $\mathcal{I} = \sigma_1$, the mirror symmetry $\mathcal{M}_y = i\sigma_3s_2$ and the rotation-mirror symmetry $\mathcal{S}_{4z} = \sigma_3e^{is_3\pi/4}$. We adopt the parameters fitted from the ARPES measurements\cite{PhysRevLett.119.267001} as $1/2m = 1375\, \mathrm{meV \AA^2}$, $\alpha = 600\, \mathrm{meV \AA^2}$, $v_{so} = 15\, \mathrm{meV \AA}$, $\mu = 55\, \mathrm{meV}$ and the lattice constant $a_0=3.89\,\AA$\cite{ding2024sublatticedichotomymonolayerfese}. Under these parameters, $h_{eff}({\bm k})$ reproduces the low-energy bands in the monolayer FeSe, supporting two Fermi surfaces located close to each other near the M point, i.e. $(\pi, \pi)$ in the Brillouin zone, as shown in Fig.\ref{fig3}(a). According to the experimental observations\cite{ding2024sublatticedichotomymonolayerfese, to.appear}, we consider an inversion-symmetry breaking term $h_{sb} = -4 \delta t \sigma_{3}$ in the monolayer FeSe, which is the leading-order perturbation and may arise from the substrate. The nodeless $d$-wave pairing state, which actually can be viewed as the $d$-wave pairing between the nearest-neighbour Fe, takes the form $h_{d-SC} = 4\Delta_d\sigma_1is_2$. Projecting the above $d$-wave pairing onto the band basis, one can find considerable interband pairing on the Fermi surfaces especially on the Brillouin zone boundary, as presented in Fig.~\ref{fig3}(a). Notice that it is due to the interband pairing, the monolayer FeSe avoides the nodes enforced by the sign change of the order parameter in the $d$-wave pairing state; Moreover, such $d$-wave pairing together with the inversion-symmetry breaking perturbation can reproduce the sublattice dichotomy on the superconducting coherence peak observed recently\cite{ding2024sublatticedichotomymonolayerfese}.

Due to the interband feature of the $d$-wave pairing, a strong JDE can be expected in the monolayer FeSe/STO. We numerically simulate the JJ constructed by the monolayer FeSe/STO with an external magnetic field applied parallel to the junction region. The Zeeman term from the magnetic field is $h_{m} ({\bm k}) = E_Zs_{2}$, where $E_Z = g\mu_BB/2$ with $B$ being the strength of the magnetic field, $g$ the Land\'{e} factor and $\mu_B$ the Bohr magneton. We note that on the Brillouin zone boundary the Hamiltonian for the above FeSe system, i.e. $h_{eff} + h_{sb} + h_{d-SC} + h_{m}$, takes exactly the same form with the Hamiltonian for the 1D junction case in the previous section. By transforming the model to its corresponding tight-binding version, we calculate the JDE and summarize the results in Figs.\ref{fig3}(b)(c). As expected, a large diode efficiency up to $\gamma\approx30\%$ can be realized. Here, it is worth pointing out that usually the JDE is highly anisotropic with respect to the angle between the pairing order and the junction, in accordance with the ansiotripic $d$-wave pairing state\cite{PhysRevB.106.214524}. However, in the monolayer FeSe case we find the anisotropy of the JDE is weak and the efficiency is always significant, due to its full-gap and interband pairing nature (details in SM\cite{SuppMat}). We also simulate the JDE in the system assuming the $d$-wave pairing coexists with the $s$-wave pairing with $h_{s-SC} = 4\Delta_s is_2$ corresponding to the $s$-wave pairing between the next-nearest-neighbour Fe, and show the results in Figs.\ref{fig3}(d). As shown, the $s$-wave component weakens the JDE; Especially in the $s$-wave dominate condition, the diode efficiency turns out to be vanishing small.

In the above, we have focused on the nodeless $d$-wave pairing. In the SM\cite{SuppMat}, we simulate the JDE in the $\eta$ pairing state and find that a diode efficiency up to $12\%$ can be achieved.

\textit{Discussion and summary.}
Besides the monolayer FeSe, our theory can also be applied to the heavy-fermion superconductor \ce{CeRh_2As_2}, whose lattice respects the same space group with FeSe. Recent experiments show that an external magnetic field can lead to a first-order phase transition in \ce{CeRh_2As_2}\cite{doi:10.1126/science.abe7518}, with the pairing order driven from even parity to odd parity according to the theoretical analysis\cite{PhysRevB.105.L020505,PhysRevB.106.L100504}. Actually, the odd-parity pairing is similar to the $\eta$ pairing in the monolayer FeSe\cite{PhysRevB.110.L140505,PhysRevB.109.024502}, which has strong interband pairing component on the Fermi surfaces. Therefore, large supercurrent diode effect can be expected in the high-field phase in \ce{CeRh_2As_2} if inversion symmetry breaking perturbations are introduced.

In summary, we reveal that in multiband superconductors the interband pairing can greatly enhance the supercurrent diode effect. Even in the weak spin-orbit coupling condition, the diode efficiency can be considerable. We apply the analysis to the monolayer FeSe/STO where signatures for interband pairing and inversion-symmetry breaking are identified in recent experiments\cite{ding2024sublatticedichotomymonolayerfese}, and show that the monolayer FeSe/STO is a potential high-temperature platform realizing significant supercurrent diode effect. In turn, we also suggest that the measurements of the supercurrent diode effect can also provide essential information on the pairing order in the monolayer FeSe/STO. Moreover, as the JDE share some similarities with the superconducting diode effect in mechanism, one can also expect strong superconducting diode effect in the monolayer FeSe/STO, which deserves further study in the future.

\textit{Acknowledgment.} This work is supported by the Ministry of Science and Technology (Grant No. 2022YFA1403900), the National Natural Science Foundation of China (Grant No. NSFC-12304163), the New Cornerstone Investigator Program, and the Beijing Institute of Technology Research Fund Program for Young Scholars.

\nocite{*}
\bibliography{references}

\end{document}